\documentclass[12pt]{article}

\usepackage[...]{youngtab}
\usepackage{a4}
\usepackage{amsfonts}
\usepackage{amssymb}
\usepackage{lscape}
\usepackage{cite}
\usepackage{epsfig}
\usepackage{multirow}
\usepackage[all]{xy}
\usepackage{amsmath}
\usepackage{amssymb}
\newcommand{\beq}{\begin{equation}}
\newcommand{\eeq}{\end{equation}}
\newcommand{\beqn}{\begin{eqnarray}}
\newcommand{\eeqn}{\end{eqnarray}}

\newcommand{\tr}{{\rm tr}}

\newcommand{\be}{\begin{equation}}
\newcommand{\ee}{\end{equation}}
\newcommand{\ba}{\begin{eqnarray}}
\newcommand{\ea}{\end{eqnarray}}
\newcommand{\bdm}{\begin{displaymath}}
\newcommand{\edm}{\end{displaymath}}

\newcommand{\ie}{{\it i.e.\ }}
\newcommand{\eg}{{\it e.g.\ }}

\DeclareMathAlphabet{\mathpzc}{OT1}{pzc}{m}{it}
\def\bea{\begin{eqnarray}}
\def\eea{\end{eqnarray}}
\def\beas{\begin{eqnarray*}}
\def\eeas{\end{eqnarray*}}
\def\sla{\raise.15ex\hbox{$/$}\kern-.57em}

\def\bea{\begin{eqnarray}}
\def\eea{\end{eqnarray}}

\def\tr{{\rm tr}}
\def\sla{\raise.15ex\hbox{$/$}\kern-.57em}
\def\ie{{\it i.e.}~}
\def\eg{{\it e.g.}~}

\begin{document}
\begin{titlepage}
\begin{flushright}
{ROM2F/2010/10}\\
\end{flushright}
%
%
\begin{center}
{\Large\bf Deforming SW curve}  \\
\end{center}
\begin{center}
{\bf  Rubik Poghossian}\\
{\sl  I.N.F.N.,  Universit\`a di Roma Tor Vergata\\
Via della Ricerca Scientifica, I-00133 Roma, Italy}\\
and \\
{\sl Yerevan Physics Institute\\
Alikhanian Br. 2, 0036 Yerevan, Armenia}
\\
e-mail: poghosyan@roma2.infn.it
\end{center}
\vskip 1.0cm
\begin{center}
{\large \bf Abstract}
\end{center}
A system of Bethe-Ansatz type equations, which specify a unique array of Young tableau responsible 
for the leading contribution to the Nekrasov partition function in the  $\epsilon_2\rightarrow 0$ limit is derived. 
It is shown that the prepotential with generic $\epsilon_1$ is directly related to the (rescaled by $\epsilon_1$) number of 
total boxes of these Young tableau. Moreover, all the expectation values of the chiral fields $\langle \tr \phi^J \rangle $ 
are simple symmetric functions of their column lengths. An entire function whose zeros are determined by the column lengths
is introduced. It is shown that this function satisfies a functional equation, closely resembling Baxter's equation 
in 2d integrable models. This functional relation directly leads to a nice generalization of the equation defining Seiberg-Witten curve.

\vfill

\end{titlepage}

\section*{Introduction}

The idea of considering ${\cal N}=2$ SYM theories in a specific presently commonly known as $\Omega$-
background is proven to be extremely fruitful. The general $\Omega$-
background is characterized by two parameters $\epsilon_1$, $\epsilon_2$ introduced in \cite{MNekShat, LosNekShat} 
to regularize the integrals over moduli space of instantons. In \cite{Nek} Nekrasov has shown how the partition function in this background 
is related to the Seiberg-Witten prepotential. In the same paper he performed explicit calculation of the prepotential up to 5 instantons 
choosing $h=\epsilon_1=-\epsilon_2$ and demonstrated that at vanishing $h$ one exactly recovers the results 
extracted from the Seiberg-Witten curve. In \cite{FP} a combinatorial formula which allows one to calculate the Nekrasov partition function 
for generic $\epsilon_1$, $\epsilon_2$ up to desired order was proposed. The partition function is represented as a sum over 
arrays of Young tableau with total number of boxes equal to the number of instantons. 

The partition function with generic $\epsilon_1$, $\epsilon_2$ is essential also 
from the point of view of the recently established AGT duality \cite{AGT} relating this partition function to the conformal blocks in 
2d Conformal Field Theory. In a parallel very interesting development Nekrasov and Shatashvily \cite{NekShat} show that
in the case when $\epsilon_2=0$ the prepotential is closely related to the quantum integrable many body systems. This case is also the 
main subject of consideration of the present paper. Note one more point which to my opinion makes the investigation 
of $\epsilon_2=0$ case even more interesting: namely, due to above mentioned AGT relation this should be related to the quasi-classical 
($c\rightarrow\infty$) limit of conformal blocks and hence to the classical Liouville field theory.  

The paper is organized as follows:
In section 1 generalizing the idea of Nekrasov and Okounkov \cite{NekOkoun} to the case with finite $\epsilon_1$ a system 
of Bethe-Ansatz type equations determining the shape of a unique, most relevant array of Young tableau is derived.

In section 2 an entire function denoted by $Y(z)$ whose zeros are determined through the column length of the 
Young tableau is introduced. It is shown that the above mentioned Bethe-Ansatz type equations are equivalent to a certain 
functional relation for the function $Y$ very much resembling the Baxter equation for the 2 d integrable models.  

In section 3  this functional relation is represented in a form which is a direct generalization of the algebraic equation defining 
the Seiberg-Witten curve \cite{SW}. The relation of the generalized Seiberg-Witten "curve" to the prepotential and the expectation
 values of the chiral fields $\langle \tr \,\phi^J \rangle $ are investigated. It is shown that they are given by  contour 
integrals which are close analogues of the period integrals of the non-deformed case.   
 It is emphasised that the prepotential is given  by the number of 
total boxes of the limiting array of Young tableau and that the expectation values of the chiral fields $\langle \tr \,\phi^J \rangle $ 
are simple symmetric functions of their column lengths.

The section 4 is devoted to the simplest case with gauge group $U(1)$ without matter hypermultiplets. In this case the 
functional equation and the generalized SW curve equation are solved in analytically in terms of the (generalized) hyper-geometric 
function $_0F_1$.

In appendix the solution of Bethe-Ansatz type equations of section 1 in few orders of instanton expansion is presented.

\section{Prepotential in the limit $\epsilon_2\rightarrow 0$}
In this section I derive saddle point equations which determine the $\epsilon_2=0$ limit of the deformed prepotential 
\bea 
W(a,m,\epsilon_1,q)=\lim_{\epsilon_2\rightarrow 0}  \log Z_{inst}(a,m,\epsilon_1,\epsilon_2,q),
\label{Prep}
\eea 
where $a$ collectively denotes all VEV's of the gauge multiplet  and $m$ the masses of possible extra matter hypermultiplets.
I will make use both equivalent representations of the Partition function as contour integral \cite{LosNekShat} 
and as sum over Young tableau \cite{Nek,FP}. The integral representation for the $k$ instanton contribution in the case of $U(N)$ gauge group 
and $f$ ($f\le 4$) fundamental hyper-multiplets \cite{Nek} reads
\be 
Z_k=\int \prod_{I=1}^{k} \frac{dx_I}{2\pi i} \chi_k(x_I)
\label{CountInt}
\ee
where
\bea
\chi_k(x_I)=\frac{1}{k!} {\prod_{I,J=1}^{k}}' \frac{(x_I-x_J) (x_I-x_J+\epsilon_1+\epsilon_2)}{(x_I-x_J+\epsilon_1)(x_I-x_J+\epsilon_2)}
\nonumber \\
\times\prod_{I=1}^k\frac{\prod_{a=1}^{f}(x_I+m_\ell)}{\prod_{u=1}^N(x_I-a_u)(-x_I+a_u-\epsilon_1-\epsilon_2)}
\label{integrand}
\eea
where the prime on the product symbol indicates that the diagonal $i=j$ factors $(x_i-x_i)$ should be omitted.
The instanton part of the partition function is
\bea
Z_{inst}=1+\sum_{k=1}^{\infty} Z_k q^k
\label{Zinst}
\eea
Following the ideology of \cite{NekOkoun} it is natural to expect that in the limit $\epsilon_2\rightarrow 0$ of our interest the main 
contribution to the  (\ref{Zinst}) will be dominated by certain pole in the integrand of (\ref{CountInt}) with large $k\sim 1/\epsilon_2$. 
It is possible to show that the poles which contribute to the integral (\ref{CountInt}) are in one to one correspondence (up to permutations of the variables $x_i$) 
with the arrays of $N$ Young tableau $Y_1,\ldots,Y_N$ with total number of boxes equal $k$. These are the same Young tableau which appear 
in already mentioned alternative combinatorial representation constructed in \cite{Nek,FP}. It is convenient to arrange the variables $x_I$ (in some fixed order) 
over the $k$ boxes. The specific values assumed by the variables $x_I$ at a pole are nothing but the eigenvalues of the instanton group $U(k)$. 
The rule, how to assign values to $x_I$ for given set of Young tableau is simple: assign to the $N$ corner boxes the expectation values 
$a_1,\ldots,a_N$, increase the value by $\epsilon_1$ ($\epsilon_2$) each time when passing to the next box in horizontal (vertical) direction. 
Thus the entry of the box $s=(i,j)$, $s\in Y_u$ would be 
\be 
x_{u,i,j}=a_u+(i-1)\epsilon_1+(j-1)\epsilon_2
\label{phases}
\ee  
Let us estimate $\log \,(\chi_k q^k)$ for a very large $k\sim 1/\epsilon_2$:   
\bea
\log (\chi_k q^k) &\sim & k\log \,q+\epsilon_2 \sum_{I,J=1}^k\left(\frac{1}{x_I-x_J+\epsilon_1}-\frac{1}{x_I-x_J}\right)\nonumber\\
&-&\sum_{I=1}^k\left(\sum_{u=1}^N \log\left( (x_I-a_u)(-x_I+a_u-\epsilon_1) \right)-\sum_{\ell=1}^f \log (x_I+m_\ell )\right)\,\,\,
\eea
The next step is to evaluate this expression at the values (\ref{phases}) replacing the discrete sums over those indices  
which are multiplied by small quantity $\epsilon_2$ (\ie the indices of type $j$ in 
eq. (\ref{phases})) by integrals. The assumption made here is very natural, the number 
of boxes in vertical ($\epsilon_2$) direction is very large, but this number multiplied by $\epsilon_2$ is expected to be finite and will be denoted as 
$\lambda_{u,i}$. The calculation is elementary and leads to the conclusion that $\log (\chi_k q^k) \sim {\cal H}/\epsilon_2$, where 
(below the indices $u, v\in1,\ldots ,N$; $\ell\in1,\ldots ,f$; $i,j\in1,2,\ldots)$:  
\bea
{\cal H}(x_u,i|\epsilon_1)=
\sum_{u,i;v,j} [-G(x_{u,i}-x_{v,j}+\epsilon_1)+G(x_{u,i}-x_{v,j}^0+\epsilon_1)\quad\quad\quad\quad\quad\quad\nonumber\\
+G(x_{u,i}^0-x_{v,j}+\epsilon_1)-G(x_{u,i}^0-x_{v,j}^0+\epsilon_1)]\nonumber\\
+\sum_{u,i;v}  [-G(x_{u,i}-a_v)+G(x_{u,i}^0-a_v)
-G(x_{u,i}-a_v+\epsilon_1)+G(x_{u,i}^0-a_v+\epsilon_1)]\nonumber\\
+\sum_{u,i;\ell}  [G(x_{u,i}+m_\ell)-G(x_{u,i}^0+m_\ell)]
+\sum_{u,i}(x_{u,i}-(i-1)\epsilon_1-a_u)\log q\hspace{1.3cm}
\label{action}
\eea     
where
\bea
G(x)=x(\log|x|-1),
\eea
and
\bea
x_{u,i}=a_u+(i-1)\epsilon_1+\lambda_{u,i}\,;\quad
x_{u,i}^0=a_u+(i-1)\epsilon_1
\eea
It is useful to regularize the expression (\ref{action}) assuming that there is an integer $L$ such that 
the (scaled) lengths of columns $\lambda_{u,i}=0$ when $i>L$. It is a particularly nice feature of the expression (\ref{action})
that its value does not depend on the upper limit of the summation indices $i,j$ provided this upper limit is chosen to be more 
or equal to $L$. This allows one to restrict the sums up to the range $L$ and pass to the limit of infinite $L$ at the final stage. 
In fact we will see below that the column lengths, extremizing the "action" (\ref{action}) are of order  $\lambda_{u,i}\sim {\cal O}(q^i)$.
Here is the extremality conditions for  (\ref{action})
\bea
-q \prod_{v,j}^{N,L} \frac{(x_{u,i}-x_{v,j}-\epsilon_1)(x_{u,i}-x_{v,j}^0+ 
\epsilon_1)}{(x_{u,i}-x_{v,j}+\epsilon_1)(x_{u,i}-x_{v,j}^0-\epsilon_1)}\,
 \frac{\prod_{\ell=1}^f(x_{u,i}+m_\ell)}{\prod_{v=1}^N(x_{u,i}-a_v+\epsilon_1)(x_{u,i}-a_v)}=1
 \label{BA}
\eea  
which, in view of \cite{NekShat} not very surprisingly, closely resembles the Bethe-Ansatz equations 
of integrable models.

\section{The functional equation}
To investigate the system of equations (\ref{BA}) in the limit of infinitely large $L$ it is useful to introduce the 
function
\bea
Y(z)=\prod_{u-1}^N e^{\frac{z}{\epsilon_1} \psi(\frac{a_u}{\epsilon_1})}\prod_{i=1}^\infty \left(1-\frac{z}{x_{u,i}}\right)e^{z/x_{u,i}^0}
\label{Y}
\eea
where
\bea
\psi(x)=\partial_z\log\Gamma(z)
\eea
Under the assumption that the column lengths tend to zero (which is equivalent to $x_{u,i}\rightarrow x_{u,i}^0$ at large $i$) 
the product (\ref{Y}) is convergent for arbitrary complex number $z$ and defines an entire function of $z$ with zeros located at 
$x_{u,i}$. In extreme case when all column lengths are zero the product (\ref{Y}) results in the entire function
\bea
Y_0(z)=\prod_{u=1}^N\frac{\Gamma\left(\frac{a_u}{\epsilon_1}\right)}{\Gamma\left(\frac{a_u-z}{\epsilon_1}\right)},
\eea
whose zeros are located at $x_{u,i}^0$.
In view of these definitions the large $L$ limit of the eqs. (\ref{BA}) can be represented as 
\bea
-\frac{q}{\epsilon_1^{2N}}\frac{Y(x_{u,i}-\epsilon_1)}{Y(x_{u,i}+\epsilon_1)}\prod_{\ell=1}^f(x_{u,i}+m_\ell)=1
\label{ABA}
\eea 
Let's introduce the notation 
\bea
Q_f(z)=\prod_{\ell=1}^f(z+m_\ell)
\eea
and consider the function
\bea
(-1/\epsilon_1)^NP_N(z)=\frac{Y(z+\epsilon_1)+\frac{q}{\epsilon_1^{2N}}Q_f(z)Y(z-\epsilon_1)}{Y(z)}.
\eea 
It does not have singularities at finite part of the complex plane since the potential poles $z=x_{u,i}$ are cancelled
due to the equality (\ref{ABA}). The behaviour at large $z$ is also easy to estimate. Indeed at large $z$ the ratio 
\bea
\frac{Y(z+\epsilon_1)}{Y(z)}\sim \frac{Y_0(z+\epsilon_1)}{Y_0(z)}=(-z/\epsilon_1)^N+ {\cal O}(z^{N-1}).
\label{largezY}
\eea
Thus the function $P_N(z)$ is in fact an $N$-th order polynomial (provided $f\le 2N$). Taking into account 
(\ref{largezY}) we see that
\bea
P_N(z)=z^N+ {\cal O}(z^{N-1})
\eea 
 for $f=1,2,\ldots,2N-1$
and 
\bea
P_N(z)=(1+q)z^N+ {\cal O}(z^{N-1})
\eea 
for the conformal case $f=2N$. 

So we finally arrive at the following functional equation for $Y$:
\bea
Y(z+\epsilon_1)+\frac{q}{\epsilon_1^{2N}}Y(z-\epsilon_1)\prod_{\ell=1}^f(z+m_\ell)=(-1/\epsilon_1)^N P_N(z)Y(z).
\label{BaxterEq}
\eea 
which very much resembles the Baxter's $T-Q$ equation well known  in the context of 2d integrable statistical systems.  

\section{Deformed Seiberg-Witten curve}
It is useful to introduce the notation 
\bea
w(z)=\frac{q}{(-\epsilon_1)^N}\frac{Y(z)}{Y(z+\epsilon_1)}
\label{DefSW}
\eea
and rewrite the functional equation in the following suggestive form
\bea
Q_f(z)w(z)w(z-\epsilon_1)-P_N(z)w(z)+q=0.
\label{DefSWcurve}
\eea
This equation supplemented with  the large $z$ asymptotic condition $w(z)=1/z^N +{\cal O}(1/z^{N+1})$ 
 (see (\ref{largezY})) generalizes the algebraic equation defining the SW 
curve to the case with finite $\epsilon_1$. The deformation is surprisingly
simple. The only difference from the standard case is the shift of one of the arguments by $\epsilon_1$.
Indeed putting $\epsilon_1=0$ in (\ref{DefSW}) and absorbing the polynomial $Q_f(z)$ by means of redefinition 
$\sqrt{Q_f(z)} w(z)\rightarrow w(z)$ one gets the standard curve equation. Of course, eq. (\ref{DefSWcurve}) no longer 
defines a curve in a usual sense and its geometric interpretation needs to be clarified yet.    
The question how the information about prepotential and the expectation values 
of chiral operators are encoded in the deformed "curve"  will be subject of the next two subsections.  

\subsection*{The prepotential}
The prepotential $W(a,m,\epsilon_1)$ defined by (\ref{Prep}) should be equal to the critical value of the "action" (\ref{action}). 
To evaluate this critical value it is more convenient first to calculate its derivative with respect to the instanton parameter $q$. 
Using (\ref{action}) and the criticality conditions (\ref{BA}) one easily gets
\bea
q\partial_q W(a,m,\epsilon_1,q)=\sum_{u,i}(x_{u,i}-(i-1)\epsilon_1-a_u)\equiv \sum_{u,i}\lambda_{u,i}
\eea  
\ie the $q\partial_q W(a,m,q)$ is simply the sum of all (rescaled) column lengths of the "critical" Young tableau! 
It is instructive to express this quantity in terms of the functions $Y(z)$, $Y_0(z)$  introduced before:
\bea
q\partial_q W(a,m,q)=\oint_{\cal C} \frac{dz}{2\pi i} z\partial_z \log \frac{Y(z)}{Y_0(z)},
\label{dW}
\eea 
where the integration contour ${\cal C}$ encloses all zeros of $Y(z)$ and $Y_0(z)$ \ie all the points
$x_{u,i}$, $x_{u,i}^0$.

\subsection*{Expectation values and the chiral ring}

The technique developed in the previous sections apply as well to the computation of the general chiral correlator
$ tr\,\phi^J$ in the gauge theory.  These correlators constitute the so called chiral ring. 
It is well known that in 4 d ${\cal N}=2$ SYM the chiral correlators    $\langle tr\, \phi^J\rangle$ can be represented as 
 \cite{Losev:2003py,NekOkoun,FFMP}, 
 \be
 \langle tr\,\phi^J\rangle= \langle tr\,\phi^J\rangle_{cl}+{1\over Z_{inst}} \sum_k \,q^k\, \int \prod_{I=1}^{k} \frac{dx_I}{2\pi i} \, \chi_k(x_I)  \, O_J(\{ x_I \})
\label{phiJint}
\ee
where the classical part of the expectation value
\bea
 \langle tr\,\phi^J\rangle_{cl}=\sum_{u=1}^Na_u^J,
\eea
$Z_{inst}$ is the instanton partition function and
\bea
 O_J( x_I )=-\sum_{I=1}^k \left[(x_I+\epsilon_1+\epsilon_2)^J-(x_I+\epsilon_1)^J-(x_I+\epsilon_2)^J-x_I^J\right].
 \eea
In the small $\epsilon_2$ limit of our interest $O_J( x_I)$ becomes
\bea
\lim_{\epsilon_2\rightarrow 0} \epsilon_2 O_J( x_I)=-\sum_{u,i}  \left[(x_{u,i}+\epsilon_1)^J-(x_{u,i}^0+\epsilon_1)^J-x_{u,i}^J+{x_{u,i}^0}^J\right].
 \eea 
Similar to the case of prepotential the saddle point approximation amounts to keeping one "critical" term in (\ref{phiJint}). The factors $1/Z_{inst}$  and 
$\chi$ cancel out and we get
\bea
\langle \tr \,\phi^J\rangle =\sum_{u=1}^Na_u^J-\sum_{u,i}  \left[(x_{u,i}+\epsilon_1)^J-(x_{u,i}^0+\epsilon_1)^J-x_{u,i}^J+{x_{u,i}^0}^J\right].
\label{phiJsum}
\eea
 Recall now the definitions of our functions  $Y(z)$, $Y_0(z)$ to rewrite the above expression in following three equivalent ways:  
\bea
\langle \tr \,\phi^J\rangle =\sum_{u=1}^Na_u^J-\oint_{\cal C} \frac{dz}{2\pi i} z^J\partial_z \left(\log \frac{Y(z-\epsilon_1)}{Y_0(z-\epsilon_1)}-\log \frac{Y(z)}{Y_0(z)} \right)\\
=\sum_{u=1}^Na_u^J-\oint_{\cal C} \frac{dz}{2\pi i} ((z+\epsilon_1)^J-z^J)\partial_z \log \frac{Y(z)}{Y_0(z)} 
\label{phiJ2}\\
=-\oint_{\cal C} \frac{dz}{2\pi i} z^J\partial_z \log \frac{Y(z-\epsilon_1)}{Y(z)}
\label{phiJ3}
\eea
Comparing (\ref{dW}) with the second representation (\ref{phiJ2}) specified to $J=2$ one gets the well known Matone relation 
\cite{Matone} between the prepotential and $\langle \tr\phi^2\rangle $  which holds for generic 
$\epsilon_1$, $\epsilon_2$ as well \cite{FFMP}. The last representation (\ref{phiJ3}) is also very interesting, it provides a
physical interpretation for the function  $w(z)$ entering in expression of the deformed SW curve (\ref{DefSW})
\bea
\langle \tr \,\phi^J\rangle =-\oint_{\cal C} \frac{dz}{2\pi i} z^J\partial_z \log w(z-\epsilon_1)
\label{SWdif}
\eea
which besides the shift by $\epsilon_1$ coincides with the standard non-deformed expression. 
Thus $\partial_z \log w(z-\epsilon_1)$ is the analogue of the SW differential. It is worth noting 
that the "classical" expectation value $a_u$ also can be represented in a similar way
\bea
a_u=-\oint_{{\cal C}_u} \frac{dz}{2\pi i} z^J\partial_z \log w(z-\epsilon_1),
\label{SWdif}
\eea
where the contour ${\cal C}_u$ encloses only the points $x_{u,i}$, $x_{u,i}^0$  with $i=1,2,\ldots $ and fixed $u$. 
Evidently this is the analogue of the A-cycle integral of the Seiberg-Witten theory.

\section{Explicit solution for $U(1)$}
The simplest case with gauge group $U(1)$ without hyper-multiplets can be analysed in full details.
The deformed SW curve (\ref{DefSW}) is now  defined as
\bea 
w(z)w(z-\epsilon_1)-(z-c)w(z)+q=0,
\label{U1defSW}
\eea 
where $c$ is a constant to be identified later. It is convenient to cast this equation into the form  
\bea 
f(x)f(x+1)-xf(x)-t=0.
\label{feq}
\eea 
The dictionary is
\bea
w(z)=-\frac{1}{\epsilon_1}f(-\frac{z-c}{\epsilon_1});\quad t=-\frac{q}{\epsilon_1^2}
\eea 
It is easy to see that the following continued fraction is a solution of (\ref{feq})
\bea
-\frac{f(x)}{t}=\frac{1}{x+\frac{t}{x+1+\frac{t}{x+2+\frac{t}{x+3+\cdots}}}}
\label{cfraction}
\eea
Gauss has investigated this continued fraction almost two centuries ago.  The answer is given by 
the ratio of (generalized) hyper-geometric functions 
\bea
-\frac{f(x)}{t}=\frac{1}{x}\,\frac{_0F_1(x+1,t)}{_0F_1(x,t)},
\eea
where the function $_0F_1$ is defined by the power series
\bea 
_0F_1(x,t)=\sum_{k=0}^{\infty}\frac{t^k}{x(x+1)\cdots (x+k-1)k!}
\eea
Thus
\bea
w(z)=\frac{q}{z-c}\,\frac{_0F_1(\frac{\epsilon_1+c-z}{\epsilon_1},-\frac{q}{\epsilon_1^2})}{_0F_1(\frac{c-z}{\epsilon_1},-\frac{q}{\epsilon_1^2})},
\eea
To fix the normalization constant $c$ notice that at large $x$ the function $_0F_1(x,t)\sim \exp(t/x)$. Hence at large $z$
\bea
w(z)\sim \frac{ 1+{\cal O}(z^2)}{z-c}
\label{wasymp}
\eea
In our case there is no singularity outside of the integration contour in  (\ref{phiJ3})  so it can be freely  deformed to a circle of a very large radius. 
Then taking into account the eq. (\ref{wasymp}) one gets convinced that 
\bea
\langle\phi\rangle \equiv a=c+\epsilon_1.
\eea 
So, the final answer is
\bea
w(z)=\frac{q}{z-a+\epsilon_1}\,
\frac{_0F_1(\frac{a-z}{\epsilon_1},-\frac{q}{\epsilon_1^2})}{_0F_1(\frac{a-z-\epsilon_1}{\epsilon_1},-\frac{q}{\epsilon_1^2})}.
\label{U1w}
\eea
For the sake of completeness let me present here also
a closed expression for the entire function $Y(z)$ entering in the definition of $w(z)$
(\ref{DefSW}) and satisfying the functional equation (\ref{BaxterEq}):
\bea 
Y(z)=\frac{\Gamma\left(\frac{a_1}{\epsilon _1}\right)}{\Gamma\left(\frac{a_1-z}{\epsilon _1}\right)}
\frac{_0F_1\left(\frac{a_1-z}{\epsilon _1},-\frac{q}{\epsilon _1{}^2}\right)}{_0F_1\left(\frac{a_1}{\epsilon _1},-\frac{q}{\epsilon _1{}^2}\right)}
\eea

It is straightforward to expand $w(z)$ around $q=0$. The result up to 4th order is:
\bea
w(z)=\frac{q}{-a+z+\epsilon _1}-\frac{q^2}{(a-z) \left(-a+z+\epsilon _1\right){}^2}-
\frac{2 q^3}{(a-z) \left(a-z-\epsilon _1\right){}^3 \left(a-z+\epsilon _1\right)}\nonumber\\-
\frac{\left(5 a-5 z+\epsilon _1\right) q^4}{(a-z)^2 \left(a-z+\epsilon _1\right)
 \left(-a+z+\epsilon _1\right){}^4 \left(a-z+2 \epsilon _1\right)}+O[q]^5\nonumber
\eea
At  $\epsilon_1=0$ this series coincides with the small q expansion of
\bea
w_0(z)=\frac{z-a-\sqrt{(z-a)^2-4q}}{2}
\label{w0}
\eea
as expected from (\ref{U1defSW}). 
We see that in the limit $\epsilon_1\rightarrow0$ the poles of (\ref{U1w}) condense around $a$ giving rise to the brunch cut 
of (\ref{w0}). This is a generic phenomenon, in the case of the gauge group $U(N)$ in small $\epsilon_1$ limit 
the familiar $N$ brunch cuts around the expectation values would emerge.

\section{Conclusions}

To summarise, a saddle point analysis of the instanton series for the Nekrasov partition function in the limit $\epsilon_2\rightarrow 0$
is performed. The criticality condition can be consistently truncated  to a finite system of Bethe-Ansatz type equations considering 
array of Young tableau with number of columns 
less or equal to $L$.  The truncated system with fixed $L$ determines all quantities up to the instanton order $q^L$. In  
large $L$ limit this system of algebraic equations is equivalent to a functional equation for an entire function whose zeros 
carry information about the lengths of columns of the Young tableau. This functional equation resembles Baxter's 
equation for 2 d integrable systems which also emerges in the context of 2d integrable field theories \cite{BazhLukZam}.   
After a simple transformation it becomes evident that this functional equation
represents a direct generalization of the algebraic equation defining the Seiberg-Witten curve. The analogue of the SW differential 
and its relation to the prepotential and chiral correlation functions is established. In particular it is shown that the derivative 
of the ($\epsilon_1$ deformed) prepotential with respect to the gauge coupling is simply the sum of all column lengths. \\
Finally, the simplest $U(1)$ case is solved analytically making use of the Gauss' method of the continued fractions.\\ 
It would be interesting to find Thermodynamic Bethe-Ansatz (TBA) like equations \cite{YY,AlZam1,AlZam2} corresponding to our 
functional  relation thus establishing direct contact with the results of the paper \cite{NekShat}.  

 \vskip 1cm

 \section*{Acknowledgments}
The author thanks A.~Belavin, R.~Flume and S.~Shatashvili for interesting discussions.
He also  would like to thank his colleagues  F.~Fucito and J.~F.~Morales
for many years of collaboration in general and for very useful discussions concerning 
this paper in particular.  \\
This work was partially supported by the European Commission FP7 Programme Marie Curie Grant Agreement
PIIF-GA-2008-221571, Italian MIUR-PRIN contract 2007-5ATT78, 
and the Institutional Partnership grant of the Humboldt Foundation of Germany.

\section*{Appendix: Instanton expansion}
In this Appendix I demonstrate how easily one gets instanton expansion directly from the Bethe-Ansatz type equation 
(\ref{BA}). 

A careful analysis shows that the structure of equations is consistent with $\lambda_{u,i}\sim {\cal O}(q^i)$. 
Having an $i$-th order solution with this property the equations (\ref{BA}) with $L=i+1$
uniquely determine not only $\lambda_{u,1},\ldots ,\lambda_{u,i}$ up to the next order $i+1$ but also in
leading order  the length of the next column $\lambda_{u,i+1}$, which automatically turns out to be of order 
${\cal O}(q^{i+1})$. In other words the $i$-th columns do not contribute
up to $(i-1)$-instanton order. Thus one can start with just $L=1$ and solve the equation step by 
step up to desired order. At each stage the problem boils down to a system of linear equations. 
$L=1$ case is simple. Here is the result
\bea
\lambda_{u,1}=\frac{-q\prod_{\ell=1}^f(a_1+m_\ell)}{\prod_{v\ne u}^{N}(a_u-a_v)(a_u-a_v+\epsilon_1)}+{\cal O}(q^2)
\eea  
Summing this expression over $u$ one gets the correct $1$-instanton prepotential (with arbitrary $\epsilon_1$ but 
$\epsilon_2=0$). In view of eq. (\ref{phiJsum}) other non-trivial checks can be performed against known 
results for $\langle \tr \, \phi^J\rangle$ (see \eg \cite{FFMP}).   
I have performed higher instanton order computations with various specific choices of $N$ and $f$ always 
finding perfect agreement with known results.
As an example below is given $2$-instanton result for the case with $U(2)$ gauge group without extra hypermultiplets: 
\bea
&&\lambda _{1,1}=\frac{-q}{2 a \epsilon_1\left(2 a+\epsilon _1\right)}-
\frac{\left(-8 a^5+4 a^4 \epsilon _1+22 a^3 \epsilon _1^2+3 a^2 \epsilon _1^3+3 a \epsilon _1^4+
\epsilon _1^5\right) q^2}{8 a^3 \epsilon _1^3 \left(2 a-\epsilon _1\right){}^2 \left(a+\epsilon _1\right) 
\left(2 a+\epsilon _1\right){}^3}+{\cal O}(q^3)\nonumber\\
&&\lambda _{1,2}=\frac{-q^2}{8 a \epsilon _1^3 \left(a+\epsilon _1\right) \left(2 a+\epsilon _1\right){}^2}+{\cal O}(q^3)\nonumber\\
&&\lambda _{2,1}=\frac{-q}{2 a \epsilon_1\left(2 a-\epsilon _1\right)}+
\frac{\left(8 a^5+4 a^4 \epsilon _1-22 a^3 \epsilon _1^2+3 a^2 \epsilon _1^3-3 a \epsilon _1^4+
\epsilon _1^5\right) q^2}{8 a^3 \left(a-\epsilon _1\right) \left(2 a-\epsilon _1\right){}^3 \epsilon _1^3 
\left(2 a+\epsilon _1\right){}^2}+{\cal O}(q^3)\nonumber\\
&&\lambda _{2,2}=-\frac{q^2}{8 a \left(a-\epsilon _1\right) \epsilon _1^3 \left(-2 a+\epsilon _1\right){}^2}+{\cal O}(q^3)
\eea

\bibliographystyle{abe}

\end{document}